\documentclass[twocolumn,aps,superscriptaddress]{revtex4}
\usepackage{amssymb}
\usepackage{amsmath,bm}
\usepackage{graphicx}
\usepackage[normalem]{ulem}
\usepackage[dvips]{color}
\usepackage{mhchem} % \ce{^{227}_{90}Th}
\renewcommand\sout{\bgroup \color{red} \ULdepth=-.5ex \ULset}

\hfuzz=\maxdimen
\begin{document}

\title{Cold dilute nuclear matter with $\alpha$-particle condensation in a generalized nonlinear relativistic mean-field model}
\author{Zhao-Wen Zhang}
\affiliation{School of Physics and Astronomy and Shanghai Key Laboratory for
Particle Physics and Cosmology, Shanghai Jiao Tong University, Shanghai 200240, China}
\author{Lie-Wen Chen\footnote{%
Corresponding author (email: lwchen$@$sjtu.edu.cn)}}
\affiliation{School of Physics and Astronomy and Shanghai Key Laboratory for
Particle Physics and Cosmology, Shanghai Jiao Tong University, Shanghai 200240, China}
\date{\today}

\begin{abstract}
We explore the thermodynamic properties of homogeneous cold (zero-temperature) nuclear matter
including nucleons and $\alpha$-particle condensation at low densities by
using a generalized nonlinear relativistic mean-field (gNL-RMF) model.
In the gNL-RMF model, the $\alpha$-particle is included as explicit degree
of freedom and treated as point-like particle with its interaction described
by meson exchanges and the in-medium effects on the $\alpha$ binding energy
is described by density- and temperature-dependent energy shift with the
parameters obtained by fitting the experimental Mott density.
We find that
below the dropping density $n_{\rm{drop}}$ ($\sim 3\times 10^{-3}$ fm$^{-3}$),
the zero-temperature symmetric nuclear matter is in the state of
pure Bose-Einstein condensate (BEC) of $\alpha$ particles while
the neutron-rich nuclear matter is composed of $\alpha$-BEC and neutrons.
Above the $n_{\rm{drop}}$, the fraction of $\alpha$-BEC decreases
with density and vanishes at the transition density $n_t$ ($\sim 8\times 10^{-3}$ fm$^{-3}$).
Above the $n_t$, the nuclear matter becomes pure nucleonic matter.
Our results indicate that
the empirical parabolic law for the isospin asymmetry dependence of
nuclear matter equation of state is heavily violated by the $\alpha$-particle
condensation in the zero-temperature dilute nuclear
matter, making the conventional definition of the symmetry energy meaningless.
We investigate the symmetry energy defined under parabolic approximation
for the zero-temperature dilute nuclear matter with $\alpha$-particle
condensation, and find it is significantly enhanced compared to the case
without clusters and becomes saturated at about $7$ MeV at very low densities
($\lesssim 10^{-3}$ fm$^{-3}$).
The critical temperature
for $\alpha$-condensation in homogeneous dilute nuclear matter is also discussed.
\end{abstract}

%\pacs{21.65.-f, 21.65.Mn, 26.50.+x, 21.30.Fe}
\maketitle
\section{Introduction}
The investigation of clustering effects in nuclear matter and finite nuclei
is currently one of hot topics in nuclear physics.
It is known that the nuclear matter system can minimize its energy by forming
light clusters at very low densities~\cite{Rop09,Typ10,Typ13a,Hag14,Giu14}.
The clustering phenomenon may exist in various nuclear and astrophysical
processes or objects,
such as nuclear ground states and excited states~\cite{Ebr12,Zho13,Yan14,He14,Aym14,THSR17},
heavy-ion collisions~\cite{ChenLW03,Gai04,ZhangYX12,Rop13,Hem15,Cao19},
core-collapse supernovae~\cite{Fis14,Fur17}, and the crust of neutron
stars~\cite{Fat10,Bha10,Ava12,Rad14,Ava17}.
The light clusters such as deuteron ($d=^{2}\rm{H}$), triton ($t=^{3}\rm{H}$),
helium-3 ($h=^{3}\rm{He}$), and $\alpha$-particle ($\alpha=^{4}\rm{He}$)
will become unbound in nuclear matter when the density is larger than a critical
density, i.e., the Mott density.
Therefore, the clustering effects will become important to understand the
properties of nuclear matter, e.g., the nuclear matter equation of state (EOS),
at low densities, especially below the Mott density.
Since the deuterons and $\alpha$-particles are bosons, the Bose-Einstein
condensation of $d$ and $\alpha$-particles may occur in dilute nuclear matter
system when the temperature is lower than the corresponding critical temperature,
and the resulting Bose-Einstein
condensate (BEC) may play an important
role in understanding the properties of the dilute nuclear matter
system.

There have a larger number of works devoting to the exploration of clustering
effects in various nuclear and astrophysical systems.
Of particular interest is the $\alpha$-clustering phenomenon
due to the specially stable structure of $\alpha$-particles.
In the last decades, great efforts have been made to study
the properties of self-conjugate $4N$ nuclei (e.g., $^{12}$C and $^{16}$O)
in order to understand the condensation of $\alpha$-particles in finite
nuclei~\cite{Toh01,Ohk04,Che07,Wak07,Sch07,Fun07,Fun08}.
The formation of $\alpha$-particles in the nuclear surface
region of heavy nuclei is investigated by Typel~\cite{Typ13b,Typ14}.
In particular, a Wentzel--Kramers--Brillouin (WKB) approximation is
used to obtain the $\alpha$-particle wave function self-consistently
with the nucleon distributions in the finite nuclei
at zero temperature.
In addition, during the past decades, the $\alpha$-clustering effects have also
been widely investigated in nuclear matter and compact
stars~\cite{Rop84,Noz85,Sch90,Ste95,Bey00}.
In the Lattimer-Swesty EOS constructed by Lattimer and Swesty~\cite{Lat91}
and the Shen EOS constructed by Shen \textit{et al}.~\cite{She11},
the $\alpha$-particles are included and treated as an ideal Boltzmann gas.
Typel \textit{et al.} study the nuclear matter
including formation of light clusters up to the $\alpha$-particle
with a generalized density-dependent relativistic mean-field
(gDD-RMF) model~\cite{Typ10}.
The EOS of dilute nuclear matter including nucleons and
$\alpha$-particles at finite temperatures
is also explored by using the virial expansion~\cite{Hor06},
and later on the additional contributions from $d$, $t$ and $h$ as well as heavier
nuclei are further included by using the S-matrix method and the
quasiparticle gas model~\cite{Mal08,Hec09,Wu17}.
Ferreira and Provind\^{e}ncia~\cite{Fer12}
explore the effect of the cluster-meson
coupling constants on the dissolution density. They use theoretical and
experimental constraints to fix the cluster-meson couplings, and obtain
the relative fractions of light clusters at finite temperature.
Furthermore, Pais \textit{et al.}
investigate the effects of the cluster-meson coupling constants on the 
properties of warm stellar matter~\cite{Pai18} and asymmetric warm nuclear matter~\cite{Pai19}.
In addition, the thermodynamic stability, phase coexistence and phase
transition in dilute nuclear matter including light nuclei
are investigated in Refs.~\cite{Typ10,Ava12,Rad14,Ava17}.

Furthermore,
Mi\c{s}icu \textit{et al.}~\cite{Mis15}
investigate the behavior of boson complex scalar fields associated with
$\alpha$-particles and anti-$\alpha$-particles by using the relativistic
mean-field (RMF) method. They consider both compressed standard baryonic matter
with small admixtures of $\alpha$-particles and systems comprising $\alpha$-particles
that are gradually doped with symmetric nuclear matter, and calculate the energy of
the momentum $\bf{k} = 0$ state, which can be viewed as the BEC of
$\alpha$ matter.
Using the momentum-projected Hartree-Fock approximation,
R\"{o}pke \textit{et al.}~\cite{Rop98,Sog09,Sog10} calculate the critical
temperature of $\alpha$-condensation in dilute nuclear matter.

In our previous work~\cite{Zha17}, a generalized non-linear relativistic
mean-field (gNL-RMF) model is developed to describe the low density nuclear matter
including nucleons, $d$, $t$, $h$, and $\alpha$-particle at finite temperature.
It is found that the clustering effect may significantly influence the EOS
of nuclear matter at low densities ($\lesssim 0.02~\rm{fm}^{-3}$).
The temperature considered in Ref.~\cite{Zha17} is above $3$~MeV,
and the $\alpha$-condensation is ignored there since it cannot
occur at the considered temperature region
in the gNL-RMF model.
Given that more and more stringent constraints on the EOS of zero-temperature asymmetric
nuclear matter, especially the zero-temperature symmetry energy, at
subsaturation densities, have been obtained from
theoretical model analyses on nuclear experimental data (see, e.g., Ref.~\cite{Zha15}),
it is thus very interesting to investigate the $\alpha$-condensation and its
influence on the low density behaviors of the EOS of zero-temperature asymmetric nuclear
matter and the symmetry energy at zero temperature,
which provides the main motivation of the present work.

In this work, we explore the properties of cold (zero-temperature) dilute nuclear matter
by using the gNL-RMF model.
At zero temperature, the dilute nuclear matter is composed of
nucleons and $\alpha$-BEC. The existence of the $\alpha$-BEC
is shown to violate the empirical parabolic law for the isospin asymmetry
dependence of nuclear matter EOS and makes the conventional definition
of the symmetry energy meaningless. The symmetry energy defined under parabolic approximation
displays a significant enhancement compared to the case without clusters
and is found to be saturated at about $7$ MeV at very low densities
($\lesssim 10^{-3}$ fm$^{-3}$).

This paper is organized as follows. In Section~\ref{Model}, we introduce
the gNL-RMF model for low density nuclear matter including nucleons and
$\alpha$-particles. And then the theoretical results and discussions are
presented in Section~\ref{Result}. Finally we give a conclusion in
Section~\ref{Summary}.

\section{Theoretical framework}
\label{Model}
In the non-linear RMF (NL-RMF) model~\cite{Wal74,Bog77,Mul96,Kub97,Liu02,Hor01,Tod05,ChenLW07},
the nonlinear couplings of mesons are introduced to reproduce the ground-state
properties of finite nuclei and to modify the density dependence of the
symmetry energy.
The gNL-RMF model~\cite{Zha17}
is an extension of the NL-RMF model by including additional light nuclei
degrees of freedom, i.e., $d$, $t$, $h$, and $\alpha$-particle.
The light nuclei are treated as point-like particle
and they interact via the exchange of various effective mesons such as
isoscalar scalar ($\sigma$) and vector ($\omega$) mesons and an isovector
vector ($\rho$) meson. The in-medium effect on the binding
energy of light nuclei is described by density- and temperature-dependent
energy shift, and the parameters of binding energy shift are determined
by fitting the Mott density extracted from the experimental data~\cite{Hag12}
(see Ref.~\cite{Zha17} for the details).

As the temperature decreases, the light nuclei except the $\alpha$-particle
make decreasing contributions to the dilute nuclear matter system.
Our calculations
indicate that the contributions of $d$, $t$ and $h$ can be neglected at
extremely low temperatures, i.e., below about $1$ MeV.
The fractions of $d$, $t$ and $h$, which are defined by
$Y_{d}=2n_{d}/n_{tot}$, $Y_{t}=3n_{t}/n_{tot}$ and $Y_{h}=3n_{h}/n_{tot}$, respectively,
are smaller than about $10^{-5}$ at such extremely low temperatures.
Therefore,
the light nuclei $d$, $t$ and $h$ are not taken into account
in the present gNL-RMF model calculations for the zero-temperature dilute nuclear matter.
The Lagrangian density of the homogeneous nuclear matter system including
nucleons, $\alpha$-particles and mesons reads
\begin{eqnarray}
\mathcal{L} &=& \sum_{i=p, n} \mathcal{L}_i + \mathcal{L}_{\alpha} + \mathcal{L}_{\rm{meson}},
\label{eq:LRMF}
\end{eqnarray}%
where the nucleons ($i=p, n$) with spin $1/2$ are described by
\begin{eqnarray}
\mathcal{L}_i = \bar{\Psi}_i\left[\gamma_{\mu}iD^{\mu}_i-M^{*}_i\right]\Psi_i,
\label{eq:Lj}
\end{eqnarray}%
while the Lagrangian density of $\alpha$-particle with spin $0$ is given by
\begin{eqnarray}
\mathcal{L}_{\alpha} &=& \frac{1}{2}\left(iD^{\mu}_{\alpha}\phi_{\alpha}\right)^{*}\left(iD_{\mu\alpha}\phi_{\alpha}\right)
-\frac{1}{2}\phi^{*}_{\alpha}\left(M^{*}_{\alpha}\right)^2\phi_{\alpha}.
\label{eq:La}
\end{eqnarray}%
The covariant derivative is defined by
\begin{eqnarray}
iD^{\mu}_i=i\partial^{\mu}- g_{\omega}^{i} \omega^{\mu}
-\frac{g_{\rho}^{i}}{2}\overrightarrow{\tau}\cdot\overrightarrow{\rho}^{\mu},
\label{eq:iDj}
\end{eqnarray}%
and the effective mass of nucleon is expressed as
\begin{eqnarray}
M^{*}=m-g_{\sigma} \sigma,
\label{eq:Mj}
\end{eqnarray}%
where $g_{\sigma}$, $g_{\omega}$, and $g_{\rho}$ are coupling constants of
$\sigma$, $\omega$, and $\rho$ mesons with nucleons, respectively;
$m$ is nucleon rest mass in vacuum which is taken to be $m=939$ MeV.
It should be noted that here neutrons and protons are assumed to have the
same mass in vacuum, but for astrophysical applications of nuclear matter
EOS, the experimental masses of neutrons ($m_{n}$) and protons ($m_{p}$)
should be adopted for accuracy and this contributes a linear splitting term in the
isospin dependence of nucleon mass.
The behavior that $\alpha$-particles dissolve at higher density can be
described by introducing the in-medium effect.
The in-medium binding energy of $\alpha$-particle $B_{\alpha}$ is related
to the $\alpha$-particle effective mass $M_{\alpha}^{*}$ via
the following relation
\begin{eqnarray}
M_{\alpha}^{*}=4 m-B_{\alpha} - g^{\alpha}_{\sigma} \sigma.
\label{eq:Mjj}
\end{eqnarray}%

The meson Lagrangian densities are given by
$\mathcal{L}_{\rm{meson}} = \mathcal{L}_{\sigma} + \mathcal{L}_{\omega}  + \mathcal{L}_{\rho} + \mathcal{L}_{\omega\rho}$ with
\begin{eqnarray}
&&\mathcal{L}_{\sigma}=\frac{1}{2}\partial_{\mu}\sigma\partial^{\mu}\sigma
-\frac{1}{2}m^2_{\sigma}\sigma^2-\frac{1}{3}g_2\sigma^3-\frac{1}{4}g_3\sigma^4, \\
&&\mathcal{L}_{\omega}=-\frac{1}{4}W_{\mu\nu}W^{\mu\nu}
+\frac{1}{2}m^2_{\omega}\omega_{\mu}\omega^{\mu}+\frac{1}{4}c_3\left(\omega_{\mu}\omega^{\mu}\right)^2,\\
&&\mathcal{L}_{\rho}=-\frac{1}{4}\overrightarrow{R}_{\mu\nu}\cdot\overrightarrow{R}^{\mu\nu}
+\frac{1}{2}m^2_{\rho}\overrightarrow{\rho}_{\mu}\cdot\overrightarrow{\rho}^{\mu},\\
&&\mathcal{L}_{\omega\rho}=\Lambda_v\left(g^2_{\omega}\omega_{\mu}\omega^{\mu}\right)
\left(g^2_{\rho}\overrightarrow{\rho}_{\mu}\cdot\overrightarrow{\rho}^{\mu}\right).
\label{eq:Lmeson}
\end{eqnarray}%
where $W^{\mu\nu}$ and $\overrightarrow{R}^{\mu\nu}$ are the antisymmetric field
tensors for $\omega^{\mu}$ and $\overrightarrow{\rho}^{\mu}$, respectively.
In the RMF approach,
meson fields are treated as classical fields and the field operators are replaced
by their expectation values.
For homogeneous matter, the non-vanishing expectation values of meson fields are
$\sigma=\langle\sigma\rangle$, $\omega=\langle\omega^0\rangle$, and
$\rho=\langle\rho^3_0\rangle$.
Noting that the cluster binding energy is density dependent, one can express the equations
of motion for the meson fields as
\begin{eqnarray}
\label{eq:sEoM}
m_{\sigma}^2\sigma &+& g_2 \sigma^2+g_3 \sigma^3 = \sum_{i=p, n, \alpha} g^i_{\sigma} n_i^s , \\
\label{eq:oEoM}
m_{\omega}^2\omega&+&c_3 \omega^3 + 2\Lambda_v g^2_{\omega} g^2_{\rho} \omega \rho^2 = \sum_{i=p, n, \alpha} g^i_{\omega} n_i \nonumber \\
&-&  \frac{m_{\omega}^2}{2g_{\omega}}\left(\frac{\partial \Delta B_{\alpha}}{\partial n^{\rm{ps}}_p}+\frac{\partial \Delta B_{\alpha}}{\partial n^{\rm{ps}}_n}\right)n_{\alpha}^s, \\
\label{eq:rEoM}
m_{\rho}^2\rho &+& 2\Lambda_v g^2_{\omega} g^2_{\rho} \omega^2 \rho = \sum_{i=p, n} g^i_{\rho} I_3^i n_i \nonumber \\
&-& \frac{m_{\rho}^2}{g_{\rho}}\left(\frac{\partial \Delta B_{\alpha}}{\partial n^{\rm{ps}}_p}-\frac{\partial \Delta B_{\alpha}}{\partial n^{\rm{ps}}_n}\right)n_{\alpha}^s,
\end{eqnarray}%
where $n_i^s$ is the scalar density, $n_i$ is the vector density,
$\Delta B_{\alpha}$ represents the binding energy shift of the
$\alpha$-particle, isospin $I_3^i$ is equal to $1/2$ for $i=p$ and
$-1/2$ for $i=n$, and following Refs.~\cite{Typ10,Typ13a,Zha17},
the meson-$\alpha$ couplings are assumed to
have the following forms
\begin{eqnarray}
g^{\alpha}_{\sigma}=4 g_{\sigma},\quad
g^{\alpha}_{\omega}=4 g_{\omega}.\quad
\label{eq:coup}
\end{eqnarray}
We note that some other forms for meson-$\alpha$ couplings
are proposed in the literate~\cite{Fer12,Ava17,Pai18,Pai19}.

The in-medium cluster binding energy $B_{\alpha}=B_{\alpha}^0+\Delta B_{\alpha}$
is dependent on temperature $T$, total proton number density $n^{tot}_{p}$, and
total neutron number density $n^{tot}_{n}$ of the nuclear matter system,
where $B_{\alpha}^0$ is the binding energy for $\alpha$-particle in vacuum and
its value is $B_{\alpha}^0 = 28.29566$~MeV~\cite{Wan17}.
The total energy shift of a cluster in nuclear medium mainly includes
the self-energy shift, the Coulomb shift and the Pauli shift.
The gNL-RMF model has already contained the self-energy shift.
The Coulomb shift can be obtained from the Wigner-Seitz approximation,
and it is very small for the $\alpha$ particle considered here
and thus neglected in the present work.
The Pauli shift can be evaluated in the perturbation theory with Gaussian approaches
for $\alpha$-particle~\cite{Typ10}.
The energy shift $\Delta B_{\alpha}$ is thus from the Pauli shift and it is assumed
to have the following empirical quadratic form~\cite{Typ10}, i.e.,
\begin{eqnarray}
\Delta B_{\alpha}(n^{tot}_{p}, n^{tot}_{n}, T)= -\tilde{n}_{\alpha}\left[1+\frac{\tilde{n}_{\alpha}}{2 \tilde{n}_{\alpha}^{0}}\right]\delta B_{\alpha}(T),
\label{eq:delbind}
\end{eqnarray}%
where $\tilde{n}_{\alpha}$ stands for
\begin{eqnarray}
\tilde{n}_{\alpha}= n^{tot}_{p} +  n^{tot}_{n},
\label{eq:abb}
\end{eqnarray}%
and the density scale for $\alpha$-particle is given by
\begin{eqnarray}
\tilde{n}_{\alpha}^{0}\left(T\right) = \frac{B_{\alpha}^{0}}{\delta B_{\alpha}\left(T\right)}.
\label{eq:denscale}
\end{eqnarray}%
The temperature dependence comes from $\delta B_{\alpha}\left(T\right)$, which is
defined by~\cite{Typ10}
\begin{eqnarray}
\label{eq:thashift}
&&\delta B_{\alpha}\left(T\right)=\frac{a_{\alpha, 1}}{\left(T+a_{\alpha, 2}\right)^{3/2}}.
\end{eqnarray}%
At a certain temperature, the Mott density of the $\alpha$-particle is obtained
when the $\alpha$-particle binding energy vanishes.
In our previous work~\cite{Zha17}, the values of
$a_{\alpha , 1} = 137330$~MeV$^{5/2}\cdot\rm{fm}^{3}$ and $a_{\alpha, 2} = 10.6701$~MeV
are obtained by fitting the experimental Mott density~\cite{Hag12}, and we also use these values
in the present work.

In the above derivations, to avoid complications due to the total baryon density
dependence of the cluster binding energies in the present theoretical framework,
following the work of Typel \textit{et al}.~\cite{Typ10}, the dependence on the
total baryon density in Eq.~(\ref{eq:delbind}) is replaced by a dependence on the
pseudodensities which are defined by
\begin{eqnarray}
\label{eq:pseudo}
n^{\rm{ps}}_n=\frac{1}{2}\left[\rho_{\omega}-\rho_{\rho}\right], \quad
n^{\rm{ps}}_p =\frac{1}{2}\left[\rho_{\omega}+\rho_{\rho}\right],
\end{eqnarray}
with
\begin{eqnarray}
\label{eq:pseudoor}
\rho_{\omega}=\frac{m^2_{\omega}}{g_{\omega}}\sqrt{\omega^{\mu}\omega_{\mu}} ,\quad
\rho_{\rho}=\frac{2m^2_{\rho}}{g_{\rho}}\sqrt{\overrightarrow{\rho}^{\mu}\overrightarrow{\rho}_{\mu}}.
\end{eqnarray}%

The clusters are treated as point-like particles, and the vector and scalar densities
of the fermions ($i=p, n$) are given, respectively, by
\begin{eqnarray}
\label{eq:fden}
n_i&=&2\int \frac{d^3k}{\left(2\pi\right)^3}\left[f_i^+(k)-f_i^-(k)\right],\\
\label{eq:fsden}
n_i^s&=&2\int \frac{d^3k}{\left(2\pi\right)^3}\frac{M_i^*}{\sqrt{k^2+M_{i}^{*2}}} \nonumber \\
&& \times \left[f_i^+(k)+f_i^-(k)\right],
\end{eqnarray}%
with the occupation probability given by the Fermi-Dirac distribution, i.e.,
\begin{eqnarray}
f_{i}^{\pm}=\frac{1}{1+\exp{\left[\left(\sqrt{k^{2}+M_{i}^{*2}}\mp\nu_{i}\right)/T\right]}}.
\label{eq:fermi}
\end{eqnarray}%
The densities of the $\alpha$-particle are obtained from
\begin{eqnarray}
\label{eq:bden}
n_{\alpha}&=&\int \frac{d^3k}{\left(2\pi\right)^3}\left[b_{\alpha}^+(k)-b_{\alpha}^-(k)\right]+n_{\rm{BEC}},\\
\label{eq:bsden}
n_{\alpha}^s&=&\int \frac{d^3k}{\left(2\pi\right)^3}\frac{M_{\alpha}^*}{\sqrt{k^2+M_{\alpha}^{*2}}} \nonumber \\
&& \times \left[b_{\alpha}^+(k)+b_{\alpha}^-(k)\right]+n_{\rm{BEC}}^s,
\end{eqnarray}%
where $n_{\rm{BEC}}$ and $n_{\rm{BEC}}^s$ are the vector and scalar density of
the $\alpha$ particles in the BEC state, respectively.
It should be noted that in homogeneous and isotropic matter $n_{\rm{BEC}}$ and
$n_{\rm{BEC}}^s$ are actually identical.
The Bose-Einstein distribution gives the occupation probability
in the following form
\begin{eqnarray}
b_{\alpha}^{\pm}=\frac{1}{-1+\exp{\left[\left(\sqrt{k^{2}+M_{\alpha}^{*2}}\mp\nu_{\alpha}\right)/T\right]}}.
\label{eq:boson}
\end{eqnarray}%
For a system including nucleons and $\alpha$-particles in chemical equilibrium as we are considering in the present work, $\nu_i$ and $\nu_{\alpha}$ are the effective chemical potentials which are defined as $\nu_i = \mu_i - g^i_{\omega}\omega - g^i_{\rho} I^i_3 \rho$ for nucleons and $\nu_{\alpha} = \mu_{\alpha} - 4 g_{\omega}\omega$ for $\alpha$-particles, respectively. The chemical potential of the $\alpha$-particle is determined by
\begin{eqnarray}
\mu_{\alpha}=2\mu_n+2\mu_p.
\label{eq:mueq}
\end{eqnarray}%

The thermodynamic quantities of homogeneous matter are easily derived
from the energy-momentum tensor. The energy density is given by
\begin{eqnarray}
\epsilon &=&\sum_{i=p, n}2\int\frac{d^3k}{\left(2\pi\right)^{3}}\sqrt{k^{2}+M_i^{*2}}\left( f_{i}^{+}+f_{i}^{-}\right) \nonumber \\
&&+ \int\frac{d^3k}{\left(2\pi\right)^{3}}\sqrt{k^{2}+M_{\alpha}^{*2}}\left( b_{\alpha}^{+}+b_{\alpha}^{-}\right) + n_{\rm{BEC}} M_{\alpha}^{*} \nonumber \\
&&+ \frac{1}{2}m_{\sigma }^{2}\sigma ^{2}+ \frac{1}{3}g_{2}\sigma ^{3}+\frac{1}{4}g_{3}\sigma ^{4} \nonumber \\
&&-\frac{1}{2}m_{\omega }^{2}\omega ^{2}-\frac{1}{4}c_{3}\omega ^{4} -\frac{1}{2}m_{\rho }^{2}\rho ^{2} \nonumber \\
&&+ \sum_{i=p, n}\left(g^i_{\omega }\omega n_i + g^i_{\rho}\rho I^i_3 n_i\right)+ 4g_{\omega }\omega n_{\alpha} \nonumber \\
&&-\Lambda_v g^2_{\omega}g^2_{\rho} \omega^2 \rho^2,
\label{eq:energy}
\end{eqnarray}%
and the pressure is obtained as
\begin{eqnarray}
p &=&\frac{1}{3}\sum_{i=p, n}2\int\frac{d^3k}{\left(2\pi\right)^{3}}
\frac{k^2}{\sqrt{k^{2}+M_i^{*2}}}\left(f_{i}^{+}+f_{i}^{-}\right) \nonumber\\
 &&+ \frac{1}{3} \int\frac{d^3k}{\left(2\pi\right)^{3}}
\frac{k^2}{\sqrt{k^{2}+M_{\alpha}^{*2}}}\left(b_{\alpha}^{+}+b_{\alpha}^{-}\right) \nonumber \\
&&- \frac{1}{2}m_{\sigma }^{2}\sigma ^{2}- \frac{1}{3}g_{2}\sigma ^{3}-\frac{1}{4}g_{3}\sigma ^{4} \nonumber \\
&&+\frac{1}{2}m_{\omega }^{2}\omega ^{2}+\frac{1}{4}c_{3}\omega ^{4} +\frac{1}{2}m_{\rho }^{2}\rho ^{2} \nonumber \\
&&+\Lambda_v g^2_{\omega}g^2_{\rho} \omega^2 \rho^2,
\label{eq:pressure}
\end{eqnarray}%
It should be noted that the condensed bosons do not contribute to the pressure
but to the energy density. The entropy density is expressed as
\begin{eqnarray}
s&=&-\sum_{i=p, n}2\int\frac{d^3k}{\left(2\pi\right)^{3}} \left[f_{i}^{+}\ln f_{i}^{+}\right. \nonumber \\
&&+\left( 1-f_{i}^{+}\right) \ln \left(1-f_{i}^{+}\right) + f_{i}^{-}\ln f_{i}^{-} \nonumber \\
&&+\left.\left( 1-f_{i}^{-}\right) \ln \left(1-f_{i}^{-}\right) \right] -\int\frac{d^3k}{\left(2\pi\right)^{3}} \nonumber \\
&& \times \left[b_{\alpha}^{+}\ln b_{\alpha}^{+}-\left( 1+b_{\alpha}^{+}\right) \ln \left(1+b_{\alpha}^{+}\right)\right. \nonumber \\
&&+ \left.b_{\alpha}^{-}\ln b_{\alpha}^{-}-\left( 1+b_{\alpha}^{-}\right) \ln \left(1+b_{\alpha}^{-}\right) \right].
\label{eq:entropy}
\end{eqnarray}%
These thermodynamic quantities satisfy the Hugenholtz-van--Hove theorem, i.e.,
\begin{eqnarray}
\epsilon=Ts-p+\sum_{i=p, n, \alpha} \mu_i n_i.
\label{eq:HH}
\end{eqnarray}%
It is convenient to define the internal energy per baryon as
\begin{eqnarray}
E_{\rm{int}}=\epsilon/n_B-m,
\end{eqnarray}%
and the free energy per baryon is given by
\begin{eqnarray}
 F=E_{\rm{int}}-T\frac{s}{n_B}.
\label{eq:free}
\end{eqnarray}%

The $\alpha$-condensation cannot occur above the critical temperature.
For a system with fixed temperature, density and isospin asymmetry, the thermodynamically
favored state can be obtained by minimizing the free energy per baryon with
respect to five independent variables, i.e., $\sigma$, $\omega$, $\rho$,
$\mu_p$ and $\mu_n$.
Below the critical temperature, the $\alpha$-condensation occurs, and
the effective chemical potential of the $\alpha$-particle equals
to the effective mass of the $\alpha$-particle, leading to that
the $\mu_p$ and $\mu_n$ are not independent at a fixed density of $n_{\rm{BEC}}$
for the $\alpha$-BEC.
In this case, we can use $n_{\rm{BEC}}$ to replace one of the two variables
$\mu_p$ and $\mu_n$ to minimize the free energy per baryon.
For a uniform three-dimensional Bose gas consisting of non-interacting particles with no
apparent internal degrees of freedom, the critical temperature for
Bose-Einstein-condensation can be expressed analytically as~\cite{Hua87}
\begin{eqnarray}
T^{\rm{Ideal}}_c = \left(\frac{n_{\rm{num}}}{\zeta(3/2)}\right)^{2/3}\frac{2\pi\hbar^2}{m_{\rm{Boson}}k_{B}},
\label{eq:Tcfga}
\end{eqnarray}%
where $n_{\rm{num}}$ is the number density, $m_{\rm{Boson}}$ is the boson rest mass,
$k_{B}$ is the Boltzmann constant, and $\zeta$ is the Riemann zeta function.
It would be interesting to compare the $T^{\rm{Ideal}}_c$ to the corresponding critical
temperature obtained from the gNL-RMF model.

\section{Results and Discussion}
\label{Result}
In our previous work~\cite{Zha17}, it is found that the clustering
effects are essentially independent of the interactions among nucleons and
light nuclei in low density nuclear matter.
In the present work, therefore, we choose only one parameter set of
the NL-RMF model, namely, FSUGold~\cite{Tod05},
for the ten parameters $m_{\sigma}$, $m_{\omega}$, $m_{\rho}$, $g_{\sigma}$,
$g_{\omega}$, $g_{\rho}$, $g_{2}$, $g_{3}$, $c_{3}$ and $\Lambda_v$.

\begin{figure}[!hpbt]
\includegraphics[width=1.\linewidth]{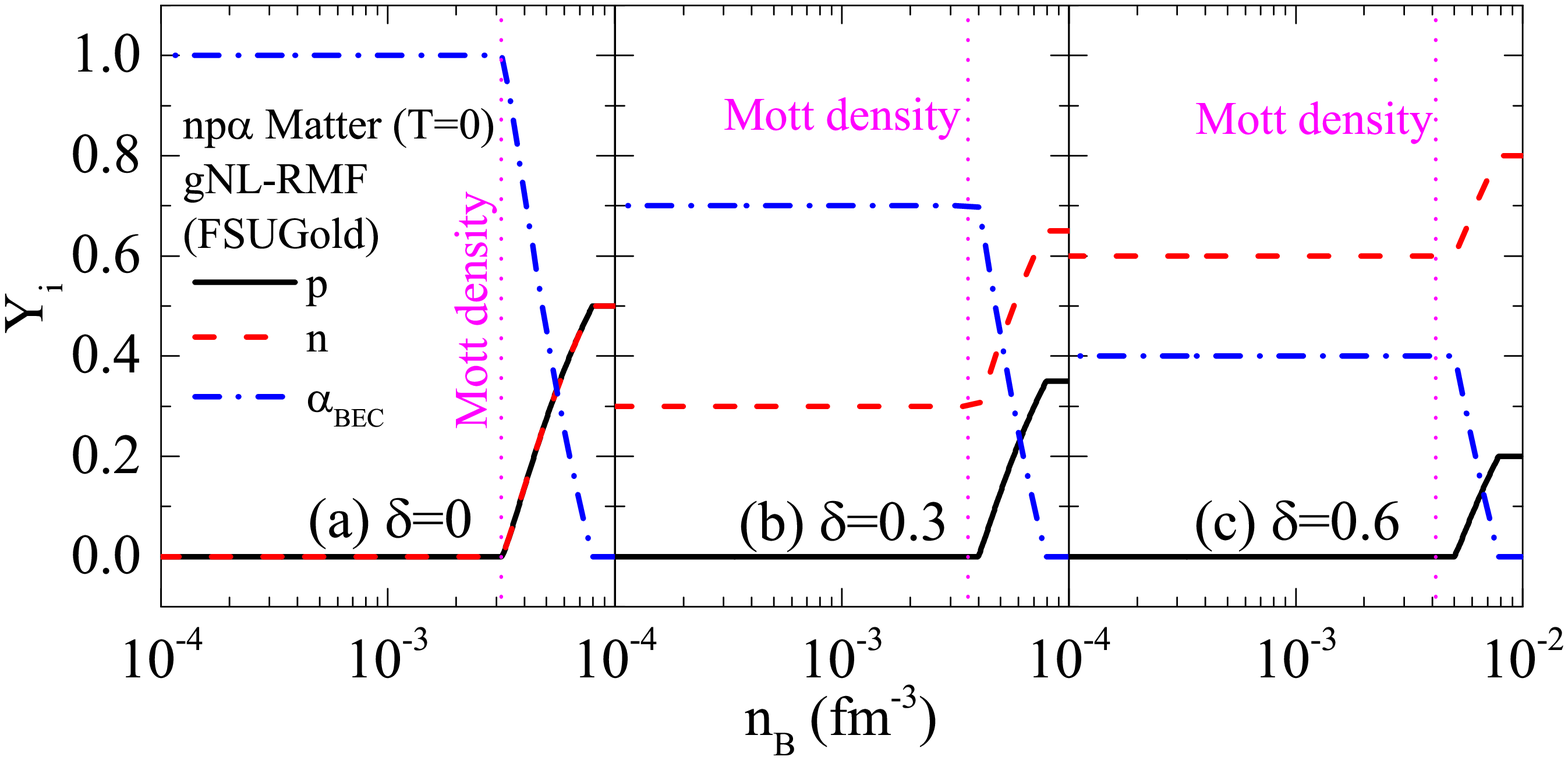}
\caption{(Color online) The fraction for nucleons and $\alpha$-particles as a function of the total baryon density $n_B$ in zero-temperature dilute nuclear matter with $\alpha$-condensation considered from the gNL-RMF model with FSUGold interaction for $\delta=0$~(a), $\delta=0.3$~(b), and $\delta=0.6$~(c). The corresponding Mott density is also indicated.}
\label{fig:Frac}
\end{figure}

Firstly, we investigate the composition of dilute nuclear matter including
$\alpha$-particles at zero temperature by using the gNL-RMF model.
At zero temperature, all of the $\alpha$-particles occupy the lowest energy
state and form $\alpha$-BEC. In addition, the $\alpha$-particle has
largest binding energy among the considered light nuclei with $A\le4$. Moreover,
in the gNL-RMF model, for the Mott density of the light nuclei
$d$, $t$, $h$ and $\alpha$ in nuclear matter at zero temperature,
the $\alpha$-particle has largest value. Therefore, only nucleons and
$\alpha$-BEC are present in the zero-temperature dilute
nuclear matter in the gNL-RMF model calculations.
Fig.~\ref{fig:Frac}
shows the fraction for nucleons and $\alpha$-particles
in the $\alpha$-BEC as a function of the total baryon density in dilute
nuclear matter system at zero temperature with isospin asymmetry $\delta = 0$, $\delta = 0.3$ and
$\delta = 0.6$, respectively.
In zero-temperature dilute nuclear matter, the fractions of neutrons, protons
and $\alpha$-particles in the $\alpha$-BEC are simply determined by the $n^{tot}_{p}$
and $n^{tot}_{n}$ ($n_B = n^{tot}_{p} + n^{tot}_{n} $) as well as the in-medium $\alpha$-particle
binding energy, and all protons are bound in the $\alpha$-BEC
when the baryon density is below a critical density
(denoted as dropping density $n_{\rm{drop}}$ above which
the $\alpha$-particle density starts to drop with density) whose value depends on
the isospin asymmetry and is slightly larger than the corresponding Mott
density ($\sim 3\times 10^{-3}$ fm$^{-3}$) at which the $\alpha$-particles
binding energy vanishes.
For the neutron-rich nuclear matter system,
therefore, the system only contains
neutrons and $\alpha$-BEC when the baryon density is below the dropping density
$n_{\rm{drop}}$ (i.e., $\sim 4\times 10^{-3}$ fm$^{-3}$ for $\delta = 0.3$ and
$\sim 5\times 10^{-3}$ fm$^{-3}$ for $\delta = 0.6$),
as observed in Figs.~\ref{fig:Frac}~(b) and (c).
It is interesting to see from Fig.~\ref{fig:Frac}~(a) that
the zero-temperature dilute symmetric nuclear matter ($\delta = 0$) becomes pure
$\alpha$-matter in the $\alpha$-BEC state when the baryon density is below
the corresponding dropping density $n_{\rm{drop}}$
(which is very close to the Mott density for $\delta = 0$).

When the baryon density is larger than
the Mott density of the $\alpha$-particle, the binding energy of
$\alpha$-particles becomes negative and so the $\alpha$-particles are
no longer in bound states,
and in this case the $\alpha$-particles may be considered as
effective resonance and continuum states.
As the density further increases, the effective resonance and continuum states
of $\alpha$-particles are expected to be continuously suppressed due to
the negative binding energy (and the fraction of nucleons gradually increases to conserve
the baryon number and isospin of the system) and
eventually disappear at a transition density $n_t$ above which
the nuclear matter becomes pure nucleonic matter.
Indeed, as expected,
one sees from Fig.~\ref{fig:Frac} that
the fraction of $\alpha$-BEC begins to decrease above the dropping density $n_{\rm{drop}}$
and then drops to zero at a critical density (i.e., the transition density $n_t$)
around $\sim 8\times 10^{-3}$ fm$^{-3}$
above which the system becomes pure nucleonic matter.
The dropping density $n_{\rm{drop}}$ is slightly larger than the corresponding Mott density
and this may be due to the interactions between the $\alpha$-particles
and nucleons (Note: the neutron density at $n_{\rm{drop}}$
is proportional to the isospin asymmetry).

\begin{figure}[!hpbt]
\includegraphics[width=1.\linewidth]{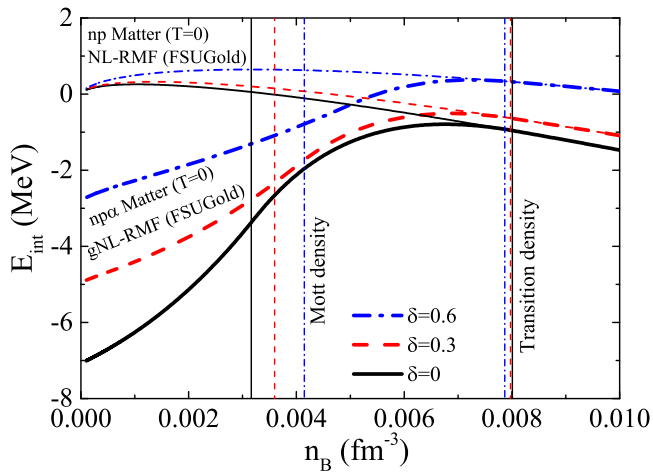}
\caption{(Color online) The internal energy per baryon $E_{\rm{int}}(n_B,\delta,T)$ as a function of the total baryon density $n_B$ in
zero-temperature dilute nuclear matter with (thick curves) and without (thin curves) considering $\alpha$-condensation from the gNL-RMF model with FSUGold interaction for $\delta=0, 0.3$ and $0.6$. The corresponding Mott density and transition density are also indicated (thin vertical lines).}
\label{fig:EBEC}
\end{figure}

Shown in Fig.~\ref{fig:EBEC} is the internal energy per baryon
as a function of the total baryon density in zero-temperature dilute nuclear
matter with and without considering $\alpha$-condensation for isospin asymmetry
$\delta=0, 0.3$ and $0.6$.
For both cases with and without considering $\alpha$-condensation, the internal
energy per baryon at a fixed baryon density increases with the isospin
asymmetry, and the increasing effect is much more pronounced in the
case with $\alpha$-condensation considered than that without considering
$\alpha$-condensation, especially at lower densities.
Generally, the symmetric nuclear matter has the minimum internal energy
per baryon.
For a fixed isospin asymmetry, it is seen that the internal energy per
baryon with and without considering $\alpha$-condensation is getting close to each
other with increasing density, and becomes identical above the transition density $n_t$,
indicating the $\alpha$ clustering
effects become weaker with increasing density.
This feature is due to the fact that above the dropping density $n_{\rm{drop}}$,
the fraction of the $\alpha$-BEC decreases with density
and the $\alpha$ particles disappear above the transition density $n_t$,
as shown in Fig.~\ref{fig:Frac}.
Compared to
the case without considering $\alpha$-condensation, the internal energy
per baryon in the case with $\alpha$-condensation considered is drastically reduced
by the formation of $\alpha$-particles.

For zero-temperature dilute symmetric nuclear matter with $\alpha$-condensation considered,
the matter actually becomes pure-$\alpha$ matter as shown
in Fig.~\ref{fig:Frac}~(a).
For pure-$\alpha$ matter, the interaction between $\alpha$-particles
is ignorable when the density tends to zero and the internal energy
per baryon is completely from the $\alpha$-BEC.
As a result, the internal energy per baryon of dilute symmetric
nuclear matter at zero temperature approaches to the negative binding energy per baryon of
the $\alpha$-particle in vacuum (i.e., $-B^0_\alpha \approx -7.1$~MeV)
when the density tends to zero, which is indeed clearly seen in Fig.~\ref{fig:EBEC}.

\begin{figure}[!hpbt]
\includegraphics[width=1.\linewidth]{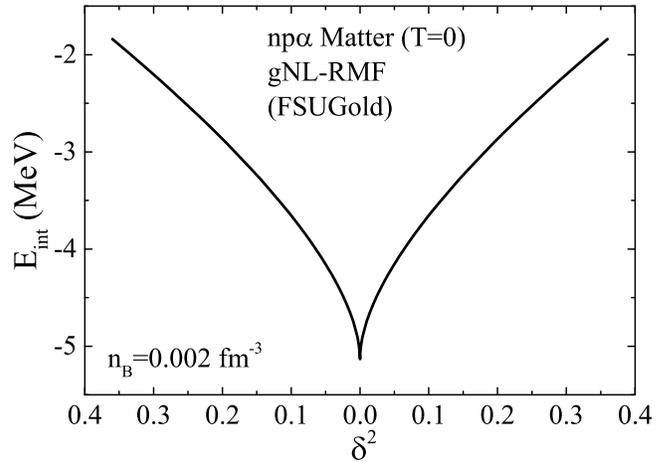}
\caption{The internal energy per baryon $E_{\rm{int}}(n_B,\delta,T)$ vs squared isospin asymmetry $\delta^2$  in zero-temperature dilute nuclear matter with $\alpha$-condensation considered at $n_B = 0.002~\rm{fm}^{-3}$ from the gNL-RMF model with FSUGold interaction.}
\label{fig:Ed}
\end{figure}

The clustering effects may break the empirical parabolic law for the
isospin asymmetry dependence of nuclear matter EOS, especially at low
temperatures~\cite{Typ10,Typ13a,Hag14,Zha17}.
It is thus interesting to check the empirical parabolic law for the dilute nuclear matter at zero temperature.
To this end,
we show in Fig.~\ref{fig:Ed} the internal energy per baryon $E_{\rm{int}}(n_B,\delta,T=0)$
as a function of the squared isospin asymmetry $\delta^2$ in dilute
nuclear matter at a representative density of $n_B = 0.002~\rm{fm}^{-3}$,
which is below the Mott density of $\alpha$-particle in symmetric and asymmetric
nuclear matter (Note: the Mott density of the $\alpha$-particle in nuclear matter
generally increase with the isospin asymmetry, as shown in Fig.~\ref{fig:Frac} and Fig.~\ref{fig:EBEC}).
As expected, the $E_{\rm{int}}(n_B,\delta,T=0)$ reaches its minimum value at $\delta=0$.
However, the curve significantly deviates from the linear relation $E_{\rm{int}} \sim \delta^2$
around $\delta=0$, indicating the violation of the empirical parabolic law.
This can be understood from the fact that the formation of $\alpha$-BEC
significantly reduces the $E_{\rm{int}}(n_B,\delta,T=0)$ as seen in Fig.~\ref{fig:EBEC},
and the system consists of only $\alpha$-BEC and neutrons with their fractions
linearly depending on the isospin asymmetry, as observed from
Fig.~\ref{fig:Frac} (see also Eqs.~(\ref{eq:nbec}) and (\ref{eq:barnd}) in the following).

We now discuss the symmetry energy of dilute nuclear matter.
Conventionally, the internal energy per baryon $E_{\rm{int}}\left(n_B, \delta, T\right)$
of isospin asymmetric nuclear matter can be expanded in powers of isospin asymmetry $\delta=\left(n^{tot}_n-n^{tot}_p\right)/\left(n^{tot}_n+n^{tot}_p\right)$ as
\begin{eqnarray}
E_{\rm{int}}\left(n_B, \delta, T\right)&=&E_{\rm{int}}\left(n_B, 0, T\right)+E_{\rm{sym}}(n_B, T)\delta^2 + \cdots,\notag\\
\label{eq:eint_sym}
\end{eqnarray}%
where the density- and temperature-dependent symmetry (internal) energy $E_{\rm{sym}}$ is defined by
\begin{eqnarray}
\label{eq:esym}
E_{\mathrm{sym}}(n_B,T)=\left.\frac{1}{2!}\frac{\partial ^2E_{\mathrm{int}}(n_B,\delta,T)}{\partial \delta^2}\right\vert_{\delta=0}.
\end{eqnarray}%
On the other hand,
under the parabolic approximation in which the higher-order expansion
coefficients (i.e., the high-order symmetry energies)
on the r.h.s. of Eq.~(\ref{eq:eint_sym}) are assumed to be small and
can be neglected, the symmetry energy can be obtained as
\begin{eqnarray}
\label{eq:esympara}
E^{\rm{para}}_{\rm{sym}}(n_B, T)&=&E_{\rm{int}}\left(n_B, \delta = 1, T\right)-E_{\rm{int}}\left(n_B,0, T\right).\notag\\
\end{eqnarray}%
Within essentially all many-body theories to date,
the parabolic approximation has been shown to be very successful for nucleonic matter
which only contains protons and neutrons,
at least for densities up to moderate values (see, e.g., Ref.~\cite{LCK08}).
Therefore, usually we have $E^{\rm{para}}_{\rm{sym}}(n_B, T) \approx E_{\mathrm{sym}}(n_B,T)$
for nucleonic matter.
For dilute nuclear matter including light nuclei,
especially at low densities ($\lesssim 0.02$~fm$^{-3}$)
and low temperatures ($\lesssim 3$~MeV), as shown in Refs.~\cite{Typ10,Typ13a,Zha17})
as well as observed from the results presented above in this work,
the higher-order expansion coefficients on the r.h.s. of Eq.~(\ref{eq:eint_sym})
could be very large and thus the expansion of Eq.~(\ref{eq:eint_sym}) may not be convergent,
leading to not very meaningful symmetry energy with the conventional definition Eq.~(\ref{eq:esym}).
Therefore,
one usually uses the $E^{\rm{para}}_{\rm{sym}}(n_B, T)$ to define the symmetry energy for
dilute nuclear matter including light nuclei
and to compare with the experimental data (see, e.g., Refs.~\cite{Typ10,Typ13a,Hag14,Zha17,Nat10}).
It should be noted that
the $E^{\rm{para}}_{\rm{sym}}(n_B, T)$ is identical to
the symmetry energy defined through the finite-difference formula
$E_{\rm{sym}}(n_B, T)=\frac{1}{2}[E_{\rm{int}}\left(n_B, \delta = 1, T\right)- 2E_{\rm{int}}\left(n_B,0, T\right) + E_{\rm{int}}\left(n_B, \delta = -1, T\right)]$~\cite{Typ10,Typ13a,Hag14,Nat10} if the mass difference
between proton and neutron (as well as between $h$ and $t$) is omitted.
Similarly, one can define the symmetry free energy and the symmetry entropy
in the same manner (see, e.g., Ref.~\cite{Zha17}).

\begin{figure}[!hpbt]
\includegraphics[width=1.\linewidth]{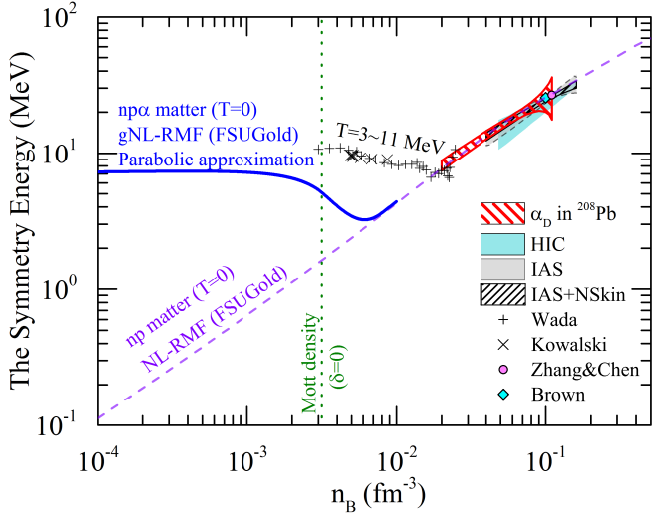}
\caption{(Color online) The symmetry energy as a function of the total baryon density $n_B$ in zero-temperature dilute nuclear matter with and without considering $\alpha$-condensation from the (g)NL-RMF model with FSUGold interaction. Some experimental constraints on the symmetry energy at zero-temperature are also included for comparison. The constraints
on $E^{\text{para}}_{\text{sym}}(n_B,T)$ in the temperature range of $T \simeq 3\sim11$~MeV extracted from heavy-ion collisions
(Wada and Kowalski)~\cite{Nat10} are included only for exploratory comparison. 
See the text for details.}
\label{fig:Esym}
\end{figure}

Figure~\ref{fig:Esym} displays the symmetry energy $E^{\rm{para}}_{\rm{sym}}(n_B, T=0)$ as a function
of the total baryon density $n_B$ in nuclear matter with and without considering
$\alpha$-condensation from the (g)NL-RMF model with FSUGold interaction.
For comparison,
we also include some experimental constraints on the symmetry energy $E_{\rm{sym}}(n_B, T=0)$, i.e.,
the constraints
from transport model analyses of mid-peripheral heavy-ion collisions of Sn
isotopes (HIC)~\cite{Tsa09},
the constraints
from the SHF analyses of isobaric analogue states (IAS) as well as combing additionally
the neutron skin ``data'' (IAS+NSkin) in Ref.~\cite{Dan14},
the constraints
from analyzing the data on the electric dipole polarizability in $^{208}$Pb
($\alpha_D$ in $^{208}$Pb)~\cite{Zha15},
the constraints
on the value of $E_{\text{sym}}$ around $2/3n_0$
($n_0 \approx 0.16$~fm$^{-3}$ is nuclear saturation density)
from binding energy difference
between heavy isotope pairs (Zhang\&Chen)~\cite{Zha13} and properties of doubly magic
nuclei (Brown)~\cite{Bro13}, and
the constraints
on $E^{\rm{para}}_{\rm{sym}}(n_B, T)$ at densities below $0.2n_0$ and temperatures in the
range $3\sim11$ MeV from the analysis of cluster formation in heavy-ion collisions
(Wada and Kowalski)~\cite{Nat10}.

It is seen from Fig.~\ref{fig:Esym} that compared to the results of pure nucleonic
matter obtained from the NL-RMF calculations with FSUGold,
the symmetry energy is drastically enhanced by including the $\alpha$-condensation
in the gNL-RMF calculations (Note: for nucleonic matter without light nuclei,
one has $E^{\rm{para}}_{\rm{sym}}(n_B, T) \approx E_{\mathrm{sym}}(n_B,T)$
in the NL-RMF calculations with FSUGold).
Very interestingly, one sees that
the $E^{\rm{para}}_{\rm{sym}}(n_B, T=0)$ in dilute nuclear matter with
$\alpha$-condensation considered is saturated at about $7$~MeV
when the baryon density is very small (less than about $n_{B} = 10^{-3}$ fm$^{-3}$).
It is constructive to examine such a saturation behavior in the low density limit
where analytic expressions might be obtained.
According to Fig.~\ref{fig:Frac}, only $\alpha$-BEC and neutrons can exist below
the dropping density $n_{\rm{drop}}$ for the neutron-rich dilute nuclear matter,
and the $\alpha$-particle number density $n_{\text{BEC}}$ in the BEC state and
neutron number density $n_{n}$ can be expressed as
\begin{eqnarray}
\label{eq:nbec}
n_{\text{BEC}}&=&n_{B}(1-\delta)/4, \\
n_{n}&=&n_{B}\delta.
\label{eq:barnd}
\end{eqnarray}%
At zero temperature, the total energy density is given by
\begin{eqnarray}
\epsilon=n_{\text{BEC}} M_{\alpha}^{*}+C n_{n}^{5/3} +n_{n} m,
\label{eq:toteneden}
\end{eqnarray}%
with $C=\frac{3}{5}\frac{\hbar^2}{2m}(3\pi^2)^{2/3}=119.1$~MeV$\cdot$fm$^2$.
At very low densities so that one has $M_{\alpha}^{*}=4m-B_{\alpha}-4g_{\sigma}\sigma\approx M_{\alpha}$,
the total energy per baryon can be expressed as
\begin{eqnarray}
E&=&\frac{1}{4}M_{\alpha} + \left(m-\frac{M_{\alpha}}{4}\right)\delta + C n_{B}^{2/3} \delta ^{5/3}.
\label{eq:totene}
\end{eqnarray}%
The second-order derivative of $E$ with respect to $\delta$ is then obtained as
\begin{eqnarray}
\frac{\partial^2 E}{\partial \delta^2}&=&\frac{10}{9} C n_{B}^{2/3} \delta^{-1/3}.
\label{eq:esym12}
\end{eqnarray}%
Therefore, in the zero-temperature case for the dilute nuclear matter
containing $\alpha$-BEC and neutrons,
the conventional definition of the symmetry energy $E_{\text{sym}}$ (i.e., Eq.~(\ref{eq:esym})) is divergent,
and so one usually uses the symmetry energy definition $E^{\rm{para}}_{\rm{sym}}$ in the parabolic
approximation (i.e., Eq.~(\ref{eq:esympara})).
From Eq.~(\ref{eq:totene}), one can obtain
\begin{eqnarray}
E^{\rm{para}}_{\rm{sym}}(n_B, T=0)&=&\left(m-\frac{M_{\alpha}}{4}\right) + C n_{B}^{2/3}.
\end{eqnarray}%
In the zero-density limit, one has
\begin{eqnarray}
E^{\rm{para}}_{\rm{sym}}(n_B\rightarrow 0, T=0)&=&\left(m-\frac{M_{\alpha}}{4}\right) \notag\\
&=&B^0_\alpha/4 \approx 7.1~\rm{MeV},
\end{eqnarray}%
and this is exactly what one has observed in Fig.~\ref{fig:Esym}.
Above about $n_{B} = 10^{-3}$ fm$^{-3}$, the $E^{\rm{para}}_{\rm{sym}}(n_B, T=0)$
in dilute nuclear matter with $\alpha$-condensation considered
decreases with density, then increases after reaching a minimum value at a density of about
$n_{B} = 6\times 10^{-3}$ fm$^{-3}$, and eventually approaches to
the $E^{\rm{para}}_{\rm{sym}}(n_B, T=0)$ for nucleonic matter above the transition
density ($\sim 8\times 10^{-3}$ fm$^{-3}$).

It is nice to see from Fig.~\ref{fig:Esym} that our present results on
the $E_{\text{sym}}(n_B,T=0)$ of nuclear matter from the gNL-RMF model with
FSUGold are in good agreement with the constraints included in the figure
for baryon density above $n_B = 0.02~\rm{fm}^{-3}$.
Unfortunately,
to the best of our knowledge, there currently have no experimental
constraints on the $E^{\text{para}}_{\text{sym}}(n_B,T=0)$ of dilute nuclear matter for
baryon density below $n_B = 0.02~\rm{fm}^{-3}$.
Our present results
provide the predictions of the $\alpha$-condensation effects on the symmetry energy in dilute
nuclear matter at zero temperature
and indicate that the $E^{\text{para}}_{\text{sym}}(n_B,T=0)$
of nuclear matter is significantly enhanced due to the $\alpha$-condensation.
We would like to point out that
the constraints
on $E^{\text{para}}_{\text{sym}}(n_B,T)$ in the density region of $n_B \simeq 0.003\sim0.03~\rm{fm}^{-3}$
and the temperature range of $T \simeq 3\sim11$~MeV extracted from heavy-ion collisions
(Wada and Kowalski)~\cite{Nat10} are included in Fig.~\ref{fig:Esym}
only for exploratory comparison.
For $n_B \simeq 0.003\sim0.03~\rm{fm}^{-3}$ and $T \simeq 3\sim11$~MeV,
there are no BEC in the nuclear matter and the clustering effects due to
the formation of $d$, $t$, $h$ and $\alpha$ significantly enhance the
$E^{\text{para}}_{\text{sym}}(n_B,T)$, leading to a reasonable agreement between
the experimental data and the gNL-RMF model predictions as shown
in Ref.~\cite{Zha17}.
In addition, one also sees a rather good agreement between the measured and 
calculated results in quantum statistical (QS) approach that takes the formation of clusters 
into account (See, e.g., Refs.~\cite{Hag14,Nat10}).
Therefore,
any experimental or model-independent information on the symmetry energy
of dilute nuclear matter at zero temperature for baryon density below
$n_B = 0.02~\rm{fm}^{-3}$ is critically useful to confirm or disconfirm our present
results based on the gNL-RMF model predictions.

\begin{figure}[tbp]
\includegraphics[width=1.\linewidth]{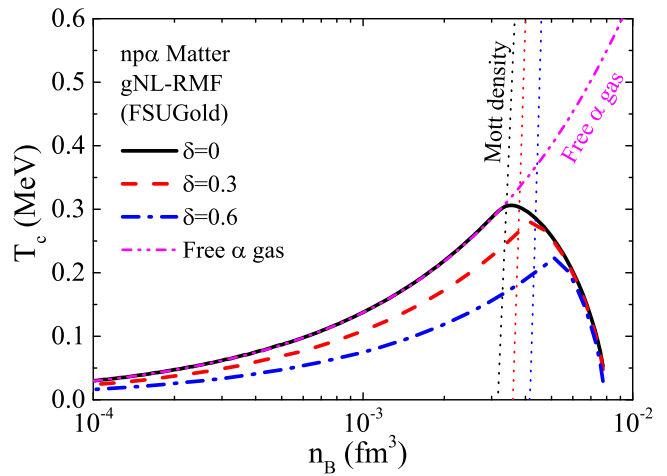}
\caption{(Color online) The critical temperature $T_c$ for $\alpha$-condensation as a function of the total baryon density $n_B$ in dilute nuclear matter obtained from the gNL-RMF model with FSUGold interaction for $\delta=0, 0.3$ and $0.6$. The corresponding Mott density is also indicated. For comparison, the critical temperature $T^{\rm{Ideal}}_c$ for free $\alpha$ gas is also included.}
\label{fig:Tc}
\end{figure}

Finally, we evaluate the critical temperature for $\alpha$-condensation
in homogeneous dilute nuclear matter within the gNL-RMF model with the FSUGold interaction.
The obtained results of the critical temperature $T_c$ vs the total
baryon density $n_B$ in homogeneous dilute nuclear matter for $\delta=0, 0.3$ and $0.6$
are shown in Fig.~\ref{fig:Tc}.
For comparison, we also include the critical temperature as a function of
the total baryon density obtained from the analytical expression $T^{\rm{Ideal}}_c$
(i.e., Eq.~(\ref{eq:Tcfga})) for free $\alpha$ gas.
It is seen that
the critical temperature in the homogeneous dilute symmetric nuclear matter
($\delta=0$) within the gNL-RMF model is almost identical to that
in the free $\alpha$ gas when the baryon density is less than
the corresponding $\alpha$-Mott-density (indicated by dotted lines in Fig.~\ref{fig:Tc}).
This is because that in the gNL-RMF model for $\delta=0$, the homogeneous dilute nuclear
matter becomes pure $\alpha$-matter and the interactions between
$\alpha$-particles are very weak at such low densities. In addition, the variation of the
$\alpha$-particle mass due to the binding energy shift in the homogeneous dilute nuclear
matter is also very small compared with the rest mass of $\alpha$-particles
in vacuum.
When the baryon density is larger than
the corresponding $\alpha$-Mott-density in the homogeneous dilute symmetric nuclear matter,
the fraction of $\alpha$ particles decreases
with density, leading to the corresponding decreasing of the $\alpha$-particle density
and thus the critical temperature, and eventually the critical temperature vanishes at
the transition density ($\sim 8\times 10^{-3}$ fm$^{-3}$),
as observed in Fig.~\ref{fig:Tc}.

Furthermore, one sees from Fig.~\ref{fig:Tc} that
the critical temperature depends on the isospin asymmetry
of the nuclear matter system, and it decreases with the isospin asymmetry.
This can be understood since for a fixed baryon density,
the number density of $\alpha$-particles decreases with the isospin
asymmetry (see, e.g., Eq.~(\ref{eq:nbec})) and thus the critical temperature
also decreases with the isospin asymmetry due to the accordingly decreasing $\alpha$-particle number density,
as shown in Eq.~(\ref{eq:Tcfga}).
In addition, for isospin asymmetry $\delta=0.3$ and $0.6$,
the corresponding critical temperature exhibits similar density dependence
as in the case of $\delta=0$, namely, it increases with density and reaches
a maximum value at a certain density,
then decreases and vanishes at the transition density
($\sim 8\times 10^{-3}$ fm$^{-3}$).
It should be noted that
the baryon density at the maximum critical temperature is
slightly larger than the corresponding $\alpha$-Mott-density,
and this may arise from the interactions between the $\alpha$-particles and nucleons,
as also observed in Fig.~\ref{fig:Frac} for the fraction of $\alpha$-BEC in
zero-temperature dilute nuclear matter.

Our results
on the critical temperature for $\alpha$-condensation in homogeneous dilute symmetric nuclear
matter below the $\alpha$-Mott-density within the gNL-RMF model are consistent with the results
from the quasiparticle gas model
obtained in Refs.~\cite{Hec09,Wu17} where the isospin dependence and the effects of
resonance and continuum states above the Mott density are not considered.
Generally speaking~\cite{Hua87}, the mean-field potentials in the gNL-RMF model can
globally influence the thermodynamic properties of the $\alpha$ matter,
but hardly affect the critical temperature for $\alpha$-condensation in
homogeneous dilute nuclear matter.
We would like to mention that,
in the present calculations, the heavier nuclei are not considered,
and only $\alpha$-particles
and nucleons are taken into account.
Including heavier nuclei, e.g., $^{56}$Fe,
may significantly influence the $\alpha$-condensation and the critical temperature,
as shown in Ref.~\cite{Wu17}.
On the other hand, as pointed out in Ref.~\cite{Wu17},
in some situations (e.g., in heavy-ion collisions or at some stages
of supernova explosions) the timescales of formation of heavier nuclei
such as $^{56}$Fe may be too long so that the light nuclei ($A\le 4$)
could be still the predominant component in the matter.
In these situations, the $\alpha$-BEC is expected to indeed occur in
the clustered matter.
In addition, it should be mentioned that the pairing effect,
which is not considered in the present work, may become important
for the Bose-condensation at higher densities
($\gtrsim 3\times 10^{-2}$ fm$^{-3}$)~\cite{Rop98}.

\section{Conclusion}
\label{Summary}
We have investigated the thermodynamic properties of homogeneous dilute
nuclear matter at zero temperature by using a generalized nonlinear
relativistic mean-field (gNL-RMF) model.
In the gNL-RMF model, the light nuclei $d$, $t$, $h$ and $\alpha$ are included
as explicit degrees of freedom and treated as point-like particles with their
interactions described by meson exchanges and the in-medium effects on their
binding energy are described by density- and temperature-dependent energy
shifts with the parameters obtained by fitting the experimental Mott densities
of the light nuclei extracted from heavy-ion collisions
at Fermi energies.

Our results have shown that in homogeneous zero-temperature dilute nuclear matter,
the binding energy of $\alpha$-particles is always larger than that of $d$, $t$ and $h$.
As a result, the $d$, $t$ and $h$ are not present in the system,
and the homogeneous zero-temperature dilute nuclear matter is composed of nucleons
and $\alpha$-particles.
In particular, when the baryon density
$n_B$ is less than the dropping density ($\sim 3\times 10^{-3}$ fm$^{-3}$),
the zero-temperature symmetric nuclear matter is found to be in the state of pure $\alpha$-BEC
and the neutron-rich nuclear matter is composed of $\alpha$-BEC and neutrons.
Above the Mott-density, the binding energy of $\alpha$-particles becomes
negative and the resonance and continuum states may appear,
which makes the fraction of the $\alpha$-particle begins to drop
at the dropping density $n_{\rm{drop}}$ and eventually vanish
at the transition density $n_t$.

In addition, we have explored the $\alpha$-condensation effects
on the symmetry energy of homogeneous dilute nuclear matter
at zero temperature.
We have shown that at zero temperature,
the existence of the $\alpha$-BEC in the homogeneous dilute nuclear matter
violates the empirical parabolic law for the isospin asymmetry dependence of
nuclear matter equation of state and makes the conventional definition of the symmetry energy meaningless.
Within the gNL-RMF model,
the symmetry energy of the zero-temperature dilute nuclear matter
defined under parabolic approximation is found to be drastically enhanced
compared to the case without considering $\alpha$-BEC,
and it becomes saturated at about $7$ MeV at very low densities ($\lesssim 10^{-3}$ fm$^{-3}$).

Finally, we have evaluated the critical temperature for $\alpha$-condensation
in the homogeneous dilute nuclear matter.
Our results indicate that the critical temperature increases with the baryon density
up to the dropping density $n_{\rm{drop}}$, then decreases and eventually vanish
at the transition density $n_t$.
In general, our results within
the gNL-RMF model gives almost identical critical temperature as that in
the free $\alpha$ gas, indicating the $\alpha$-particle interactions
are not important in the homogeneous dilute nuclear matter,
which is consistent with the result of the quasiparticle gas model.

\begin{acknowledgments}
This work was supported in part by the National Natural Science
Foundation of China under Grant No. 11625521, the Major State Basic Research
Development Program (973 Program) in China under Contract No.
2015CB856904, the Program for Professor of Special Appointment (Eastern
Scholar) at Shanghai Institutions of Higher Learning, Key Laboratory
for Particle Physics, Astrophysics and Cosmology, Ministry of
Education, China, and the Science and Technology Commission of
Shanghai Municipality (11DZ2260700).
\end{acknowledgments}

\end{document}